\documentstyle[11pt,newpasp,twoside,epsf]{article}
\begin{document}

\title{
Application of an objective method for the identification of Young Star Groupings
in spiral galaxies.
}

\author{
R. Capuzzo--Dolcetta$^1$, A. Vicari$^1$, P. Battinelli$^2$, G. Arrabito$^3$
}

\affil{$^1$Dip. di Fisica, Univ. La Sapienza, P.le A.Moro 2, 
00185 Roma (I)\\ $^2$Oss. Astron. Roma, V.le del Parco
Mellini 84, 00136 Roma (I)\\$^3$Dip. di Matem., Univ. La
Sapienza, P.le A.Moro 2, 00185 Roma (I)}


\begin{abstract}
We present some results of the application of our automatic
technique for the identification of Young Star Groupings (YSGs) in spiral
galaxies. In this report we discuss our new results for NGC 3377a and NGC 3507
together with previously published data of NGC 7217, NGC 1058 and UGC 12732.
The aim of this ongoing research is to obtain a wide sample of
homogeneous data for the study of the YSGs and their relationships with
the properties of their parent galaxies, using only objective identifications.


\end{abstract}

\section{Introduction}
Several works have pointed out the importance of the study of YSGs in
galaxies and, in particular, of the study of the relations between 
properties of the YSGs and global properties of the mother galaxy.
These relations are important for deducing properties of the
physical mechanism of star formation and, possibly, as indirect distance
estimates.
Examples of such correlations found in the literature are the scaling laws
between the largest YSG in a galaxy and the size of the galaxy itself
(Elmegreen et al. 1994, 1996), or the relationship between the total B magnitude
and color indices of the galaxy and the luminosity of the greatest YSG
(Wray et al. 1980). To check these laws and to deducing new
interesting relations, we have developed an automatic method for the
identification of the YSGs in unresolved galaxies. The method is based on a
multivariate statistical analysis (Adanti et al. 1994, Battinelli et al.
2000).

\begin{figure}
\plotfiddle{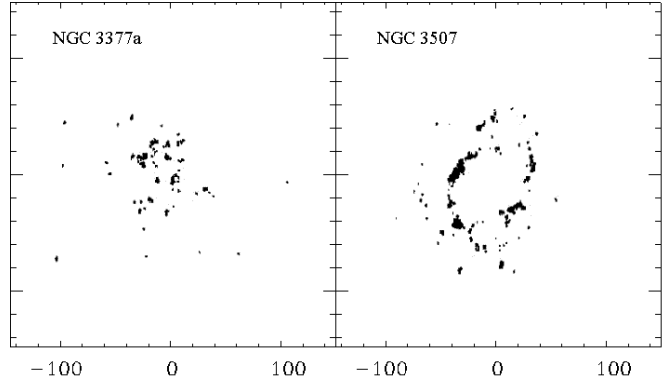}{3.0cm}{0}{50}{50}{-150}{-150}
\caption{Map of the YSGs identified in NGC 3377a (left panel) and in NGC 3507 (right panel).
Coordinates are in arcseconds with the offset at the galactic center. North is up and East is to the left}
\end{figure}

\begin{figure}
\plotfiddle{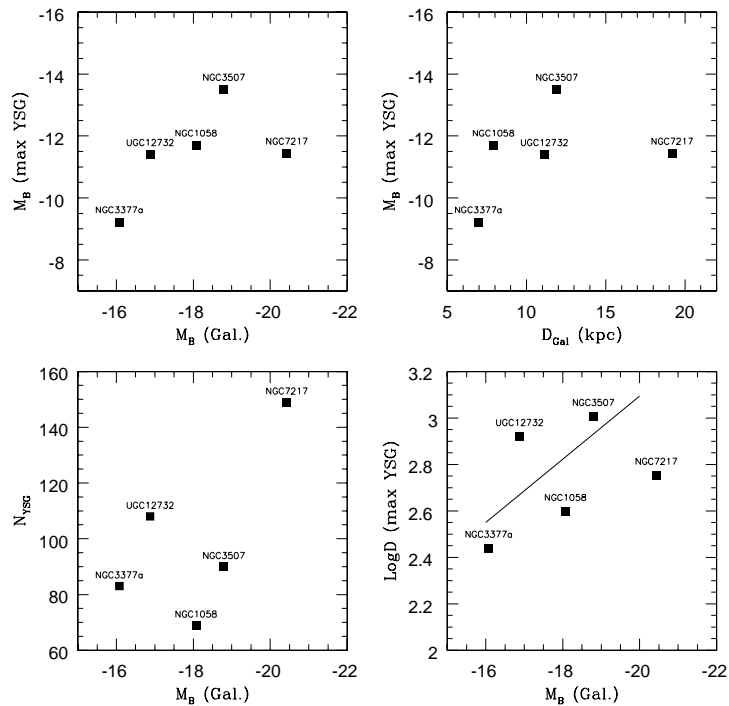}{2in}{0}{50}{50}{-150}{-90}
\caption{
Some interesting correlations between YSGs properties and parent galaxy integral
properties. M$_B$ is the absolute total B magnitude of the galaxy. The straight
line has the slope of the relationship found by Elmegreen et. al, 1994.}
\end{figure}

\section{Results and discussions}
We have applied our technique to five almost face-on spiral
galaxies, so far: NGC 7217, UGC 12732, NGC 1058, NGC 3377a and NGC 3507. Data
were taken with the APO 3.5 meter telescope. Data and results for the first
three galaxies are published in Battinelli et al. 2000, while data for the
last two galaxies have been just recently obtained and not yet published. We find
149, 108, 69, 83 and 90 YSGs respectively. For each YSG we determine position, dimension, (as average of the x and y
extents) and integrated blue magnitude. Here we report some preliminary
results: for the unpublished galaxies we give the map of the distribution of the YGSs (Figure 1).
Galaxy distances were estimated by mean of the Hubble law, with H$_0$=75
Km/s/Mpc and radial velocities from RC3 (de Vaucouleurs et al. 1991), except
for NGC 3377a, whose distance is obtained from Cepheids (Tonry et al. 1990).
The uncertainty on the contribution of  the peculiar motion to the radial
velocities is the principal source of error.
\par Table 1 resumes some of the data
displayed in Figure 2, that shows some correlations among some YSGs parameters
and properties of their mother galaxies. By the way, to draw firm conclusions
about these relationships require a larger data sample and more reliable
distance determinations.

\begin{table}
\center
\caption{
Some parameters characterizing our identified YSGs, together with data of their parent galaxies. D is the distance (see text). B$_0$$^T$ is
the total blue magnitude. LogD$_0$ is the logaritm of the galactic diameter
(all these data are from RC3). N$_{YSG}$ is the number of the YSGs identified
in this work. D$_{max}$ is the diameter of the largest YSG in the galaxy and
B$_{max}$ is its integrated blue magnitude.}
\begin{tabular}{lllllll}\tableline
	 {\sf Name}
	&{\sf D (Mpc)}
	&{\sf B$_0$$^T$}
	&{\sf LogD$_0$}
	&{\sf N$_{YSG}$}
	&{\sf D$_{max}$ (pc)}
	&{\sf B$_{max}$}
\\\tableline
	 {\sf NGC 1058}
	&{\sf 8,40}
	&{\sf 11,55}
	&{\sf 1,51}
	&{\sf 69}
	&{\sf 396}
	&{\sf 17,93}
\\
	 {\sf NGC 3377a}
	&{\sf 10,70}
	&{\sf 14,07}
	&{\sf 1,35}
	&{\sf 83}
	&{\sf 275}
	&{\sf 20,93}
\\
	 {\sf NGC 3507}
	&{\sf 12,10}
	&{\sf 11,62}
	&{\sf 1,53}
	&{\sf 90}
	&{\sf 1016}
	&{\sf 16,92}
\\
	 {\sf NGC 7217}
	&{\sf 15,50}
	&{\sf 10,52}
	&{\sf 1,63}
	&{\sf 149}
	&{\sf 567}
	&{\sf 19,52}
\\
	 {\sf UGC 12732}
	&{\sf 12,40}
	&{\sf 13,59}
	&{\sf 1,49}
	&{\sf 108}
	&{\sf 833}
	&{\sf 19,08}
\\\tableline
\tableline
\end{tabular}
\end{table}



\begin{references}
	\reference Adanti S., Battinelli P., Capuzzo--Dolcetta R., Hodge P.W. 1994, A\&AS, 108, 395
	\reference Battinelli, P., Capuzzo--Dolcetta, R., Hodge, P.W. 1994, Vicari A., Wyder T. K., A\&AS, 108, 395
	\reference de Vaucouleurs, G., de Vaucouleurs, A., Corwin, H.G.Jr., et al., 1991,{\it Third Reference Catalogue of Bright Galaxies}, (Berlin: Springer-Verlag)
	\reference Elmegreen, D.M., Elmegreen, B.G., Lang, C., Stephens, C.  1994, ApJ, 425,57
	\reference Elmegreen, B.G, Elmegreen, D.M., Salzer, J., Mann, H.  1996, ApJ, 467, 579
	\reference Hodge, P.W. 1986, in {\it Luminous Stars and Associations
	in Galaxies}, IAU Symp. n.116 (Dordrecht:Reidel)
	\reference Tonry J.L., Blakeslee J.P., Ajhar E.A., Dressler A.,1997, ApJ, 475, 399
	\reference Wray J.D., de Vaucouleurs G., 1980, ApJ. 85 (1), 1-8
\end{references}
\end{document}